\documentclass[twocolumn,showpacs,preprintnumbers,amsmath,amssymb]{revtex4}	
\usepackage{graphicx}
\usepackage{dcolumn}
\usepackage{bm}

\begin{document}

\title{Debye representation of dispersive focused waves}

\author{Carlos J. Zapata-Rodr\'{\i}guez}
\email{carlos.zapata@uv.es}
\affiliation{Department of Optics, University of Valencia, 46100 Burjassot, Spain.}

\date{\today}

\begin{abstract}
We report on a matrix-based diffraction integral that evaluates the focal field of any diffraction-limited axisymmetric complex system.
This diffraction formula is a generalization of the Debye integral applied to apertured focused beams, which may be accommodated to broadband problems.
Longitudinal chromatic aberration may limit the convenience of the Debye formulation and, additionally, spatial boundaries of validity around the focal point are provided.
Fresnel number is reformulated in order to guarantee that the focal region is entirely into the region of validity of the Debye approximation when the Fresnel number of the focusing geometry largely exceeds unity.
We have applied the matrix-based Debye integral to several examples.
Concretely, we present an optical system for beam focusing with strong angular dispersion and free of longitudinal chromatic aberration.
This simple formalism leaves an open door for analysis and design of focused beams with arbitrary angular dispersion.
Our results are valid for ultrashort pulsed and polychromatic incoherent sources.
\end{abstract}


\maketitle

\section{Introduction}

Interaction of high-intense ultrashort pulsed radiation with matter enhances multiphoton ionization \cite{Corkum93,Codling93}, high-harmonic generation \cite{Bartels00,Oron05}, and supercontinuum formation\cite{Wadsworth02,Dharmadhikari04}.
The above experimental demonstrations give rise to significant applications such as multiphoton excitation fluorescence imaging \cite{Konig00} and high-harmonic microscopy \cite{Cernusca98,Barad97}, where microscopic structures of transparent samples may be probed in the vicinity of the focus of a tightly focused beam.
Such broadband radiation has intrinsically a serious sensitivity to temporal dispersion and chromatic aberrations \cite{Kempe92,Kempe93}.
Consequently, spatio-temporal control of broadband light focusing is mandatory.

Some authors have demonstrated simultaneous spatial and temporal focusing of femtosecond pulses by means of wavefield division \cite{Zhu05,Zeng06}.
This technique is based on the spatial separation of the spectral components of pulses into a collection of off-axis beams, also called ``rainbow beam'', thus allowing a parallel processing, and recombining these components at the focal point of an achromatic objective lens.
The proposed arrangements incorporate diffractive gratings, frequently combined with refractive prisms, with specific dispersive behavior.
Similarly, generation of multiple spots in the back focal plane of a lens results in a notable energy dispersion, and some proposals with spatial-dispersion compensation properties have been presented elsewhere \cite{Amako02,Li05}.
Interestingly, the diffraction gratings may be appropriately substituted by Fresnel lenses for spatial dispersion compensation of incoherent white light \cite{Morris81,Lancis99}, profiting from the inherent radial symmetry of these optical elements.
A complete system design may be performed in terms of the ABCD transfer matrix theory \cite{Lancis04}.

The focal region is extensively analyzed in the literature under the assumption that the incoming spherical wave is monochromatic and the beam is tightly focused such that the Fresnel number of the focusing arrangement is high \cite{Stamnes86,Born99}.
The Fresnel number of the apertured focusing lens, $N = a^2 / \lambda f$, depends on the aperture radius, $a$, the lens focal length, $f$, and the radiation wavelength, $\lambda$.
In a general diffraction problem, the Kirchhoff diffraction theory should be used.
However, when the Fresnel number is much higher than unity, some spatial symmetries in the vicinity of focus are found, and the Debye approximation holds \cite{Collet80,Wolf81}.
By the way, the concept of Fresnel number has recently been interpreted for broadband coherent radiation \cite{Pearce02}.
However, roughly speaking, the Debye diffraction formulation has been ignored in the evaluation of the focal field of polychromatic (both ultrashort pulsed and temporally incoherent) spherical waves.
Only a few studies have taken into account the aforementioned approximation \cite{Ohara98,Gbur02,Zapata06,Zapata06b}.

In the transfer matrix theory, the Fresnel-Kirchhoff integral is expressed in terms of the Collins formula \cite{Collins70}, which proves to be a powerful tool for system analysis and design \cite{Siegman86,Yura87}.
Obviously, we may include the analysis of either monochromatic or polychromatic focal waves of complex optical systems.
A large number of focusing problems encounter the natural field symmetries about the focal point and, therefore, they are opened to be expressed in terms of the Debye representation.

The aim of this paper is to investigate the transfer-matrix formalism within the Debye approximation, and accordingly to present a simplified form of the Collins diffraction integral.
In particular, we develop the diffraction matrix formulation for broadband complex optical arrangements.
The paper is organized as follows.
In Sec.~II the basic grounds on diffraction of apertured focused waves in the Debye approximation are reviewed, adding emphasis to polychromatic waves.
In Sec.~III the ray matrix theory is introduced.
In this way, we obtain the longitudinal and angular dispersion inherent to broadband focal waves. 
Additional constraints among the matrix elements are achieved in a telecentric optical arrangement where the Debye representation exactly evaluates the focal field.
In Sec.~IV we extend the Debye representation for the evaluation of the focal field in diffraction-limited complex optical systems.
The region of validity of the Debye approximation is derived and the role of the Fresnel number to enclose the focal volume is pointed out.
In Sec.~V we analyze some examples.
Refractive and diffractive apertured singlets are studied in detail.
The dispersive behavior of the numerical aperture and the Fresnel number is shown.
We also discuss the dispersion behavior of a version adapted for focusing purposes of the optical system of Ref.~\cite{Lancis04} with nearly frequency-independent Fraunhofer patterns.
Finally, in Sec.~VI the main conclusions are outlined.

\section{Debye representation of focal waves}

\begin{figure}
\includegraphics[width=8cm]{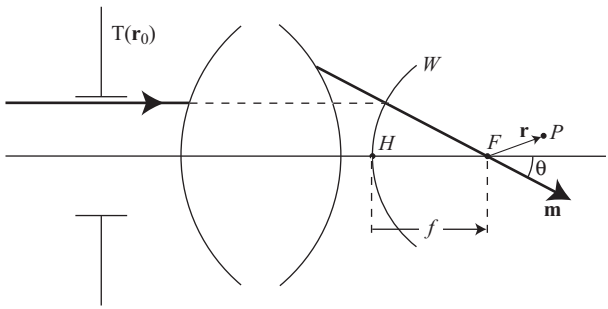}
\caption{Schematic illustration of the optical system under consideration.
 For simplicity, we have depicted the input and output optical surfaces of the focusing arrangement.}
\label{fig01}
\end{figure}

Consider a uniform plane wave incident normally onto a focusing optical system.
In front of the optical system we place an aperture, of amplitude transmittance $T(\mathbf{r}_0)$, at a transverse plane hereafter called input plane, which governs the amplitude and the extent of the emerging beam.
Consider a ray of the incident beam travelling parallel to the optical axis and passing though the aperture at a point of the input plane $\mathbf{r}_0 = (r_0,\phi_0)$ (in polar coordinates).
If the focusing system is aberration-free, the emerging ray is driven to the back focal point $F$ located at a distance $f$ from the second principal point $H$ (see Fig.~\ref{fig01}).
A reference spherical surface, $W$, of centre in the focal point $F$ and radius $f$ reproduces the wavefront of the emerging focusing wave.
When the sine condition is satisfied, the trajectories of the incident and emerging rays intersect on the reference surface.

In general, we may describe the direction of propagation of the emerging geometrical ray by means of the unitary three-dimensional (3-D) vector $\mathbf{m} = (\sin \theta \cos \varphi, \sin \theta \sin \varphi, \cos \theta)$ written in the Cartesian coordinate system.
From the geometry of Fig.~\ref{fig01} we deduce that the azimuth of the 3-D vector $\mathbf{m}$ is given as 
\begin{equation}
\varphi = \phi_0 + \pi \ .
\label{eq01}
\end{equation}
The sine condition imposes a constraint for the zenith giving
\begin{equation}
 \sin \theta = \frac{r_0}{f} \ ,
\label{eq02}
\end{equation}
where $g(\theta) = \sin \theta$ is called the ray projection function \cite{Gu00}.
In general, $g(\theta)$ gives how a ray entering the optical system is projected on the wavefront $W$, and may have different mathematical expressions other than the sine function.
For example, under the Herschel condition the ray projection function is $g(\theta) = 2 \sin(\theta / 2)$.
However, in the paraxial approximation ($\theta \ll 1$) the ray projection function is unique, $g(\theta) = \theta$. 

The focal field may be evaluated in terms of the Debye formulation, which considers the interference of plane waves with propagation directions given by the vector $\mathbf{m}$.
The amplitude of the field at a point $P$ of the focal region is then written as \cite{Born99}
\begin{equation}
 U(\mathbf{r}) = - \frac{i k}{2 \pi} \int_0^{2 \pi} \int_0^\alpha Q(\theta,\varphi) 
  \exp{\left( i k \mathbf{m} \mathbf{r} \right)} \sin \theta 
   \mathrm{d} \theta \mathrm{d} \varphi \ ,
\label{eq03}
\end{equation}
where $Q(\theta,\varphi)$ is the apodization function, $\mathbf{r}=(r \cos \phi,r \sin \phi,z)$, and $k = \omega / c$ is the wavenumber.
The numerical aperture of the focusing arrangement, $NA = \sin \alpha$, is obtained from the equation $g(\alpha) = R/f$, where $R$ is the maximum lateral extent of the aperture.
Whereas $|T|^2$ provides the ray density in the transverse input plane, the squared modulus of the apodization function $|Q|^2$ gives the ray density over the spherical reference surface $W$.
From the consideration of energy balance \cite{Gu00} 
we find a relation of the apodization function and the transmittance of the diffracting aperture
\begin{equation}
 \frac{Q(\theta,\varphi)}{f} = T\left(f g(\theta),\varphi + \pi\right) 
  \sqrt{\left|\frac{\partial_\theta g^2(\theta)}{2 \sin \theta}\right|} \ .
\label{eq04}
\end{equation}
Then in the focal volume we have a superposition of plane waves of amplitude distribution $Q$ and propagating in directions given by $\mathbf{m}$.

When the incident plane wave is polychromatic, we may use Eq.~(\ref{eq03}) to evaluate the amplitude distribution in the focal region for the frequencies constituting the field spectral range. 
In general, an optical system is unable to focus the field of different frequencies in the same focal point, being distributed along the optical axis (longitudinal chromatic aberration).
Compensation of this spatial dispersion is available, though the focus position is not rigorously independent upon frequency. 

Let us write the Debye diffraction integral for a frequency $\omega$ as follows
\begin{eqnarray}
 U_\omega(\mathbf{r}) = \frac{\omega}{i 2 \pi c} \int_{-\pi}^\pi \int_0^\alpha 
  Q(\theta,\phi_0 + \pi) \nonumber \\
  \exp{\left[ i \frac{2 \omega}{c} z_0 \sin^2\left(\frac{\theta}{2}\right) \right]} 
  \exp{\left[ - i \frac{2 \omega}{c} z \sin^2\left(\frac{\theta}{2}\right) \right]} \nonumber \\
  \exp{\left[ - i \frac{\omega}{c} r \sin \theta \cos(\phi - \phi_0) \right]}
   \sin \theta \mathrm{d} \theta \mathrm{d} \phi_0 \ .
\label{eq05}
\end{eqnarray}
In the previous equation we have dropped a factor $\exp{(i k z)}$, and we use the trigonometric formula 
\begin{equation}
 \cos \theta - 1 = - 2 \sin^2 \left(\theta / 2 \right) \ .
\label{eq06}
\end{equation}
Also, we have used Eq.~(\ref{eq01}) in order to integrate in the angular variable $\phi_0$.
If $F_0$ represents the focal point corresponding to a given reference frequency $\omega_0$, the parameter $z_0(\omega)$ stands for the axial distance from $F_0$ to the geometrical focus for the frequency under consideration.

Moreover, the position of the second principal point $H$ is also spatially dispersed, thus affecting to the value of the focal length.
Even in a case where the focal point is strictly independent of the frequency, a dispersive focal length $f(\omega)$ caused by an axial shift of $H$ would alter the angular distribution $\theta(\omega)$ of plane waves in the focal region, as deduced from Eq.~(\ref{eq02}).
This phenomenon is denominated as angular dispersion, and is responsible of the dispersive character of the numerical aperture.
Although spatial and angular dispersion are commonly neglected in achromatic objectives, we may design some optical arrangements with distinctive dispersive behavior.
Finally, the ray projection function $g(\theta)$ may also vary with the wavelength, though it will be neglected in the present paper.

Previously to the use of the Debye diffraction formula of Eq.~(\ref{eq05}) for broadband beams, the (spatial and angular) dispersion attributes of the focusing system should be given, represented in the parameters $z_0(\omega)$ and $f(\omega)$, together with the ray projection function.
According to the paraxial approximation, if the numerical aperture of the focused wave is sufficiently small, $\tan \theta$ and $\sin \theta$ may be replaced by $\theta$, and the ray projection function is reduced to $g(\theta) = \theta$.
Performing an additional geometrical transformation $\theta = r_0 / f$ in the diffraction integral, based on the paraxial approximation of Eq.~(\ref{eq02}), the Debye diffraction integral may be rewritten as
\begin{eqnarray}
 U_\omega(\mathbf{r}) = \frac{\omega}{i 2 \pi c} \frac{1}{f} 
  \int_{-\pi}^\pi \int_0^R T(r_0,\phi_0)
  \exp{\left( i \frac{\omega}{2 c} \frac{z_0}{f^2} r_0^2 \right)} \nonumber \\
  \exp{\left( - i \frac{\omega}{2 c} \frac{z}{f^2} r_0^2 \right)} 
  \exp{\left[ - i \frac{\omega}{c} \frac{r r_0}{f} \cos(\phi - \phi_0) \right]}
   r_0 \mathrm{d} r_0 \mathrm{d} \phi_0 \ ,
\label{eq07}
\end{eqnarray}
where we have used Eq.~(\ref{eq04}).
In this case, the 2-D integration is performed in the input plane in lieu of the reference spherical surface.
Importantly, the focal wavefield given in the previous integral equation satisfies the parabolic wave equation
\begin{equation}
 [\partial_r^2 + r^{-1} \partial_r + r^{-2}\partial_\phi^2 + 2 i (\omega / c) \partial_z] U_\omega(\boldsymbol{r}) = 0 \ .
\label{eq08}
\end{equation}

\section{Transfer matrix method}

The dispersive behaviour of an aberration-free ($z_0$ independent of $r_0$ for a given frequency) optical system, which is characterized by the focal length $f(\omega)$ and the longitudinal dispersion $z_0(\omega)$, is analyzed hereafter for focused beams satisfying the paraxial condition ($\theta \ll 1$).
Under the paraxial regime, beam propagation in the optical system may be represented with the $ABCD$ matrix method.
Thus we follow a procedure to obtain the values of $f$ and $z_0$ by means of the elements of the $ABCD$ matrix of the optical system.
The transfer matrix describes both the incident and emerging beams in basis of geometrical rays.

Consider a ray incident onto the optical systems that, at the input plane, propagates at a height $r_0$ with an angle $\theta_0$ with respect to the $z$-axis.
If the ray traverses an optical system described with a transfer matrix 
\begin{equation}
M_0 = 
 \left[ \begin{array}{ccc}
   A_0 & B_0 \\
   C_0 & D_0
 \end{array} \right] \ ,
\label{eq09}
\end{equation}
the emerging ray exits with a height $r$ and an angle $\theta$ given by
\begin{equation}
 \left[ \begin{array}{ccc}
   r  \\
   \theta 
 \end{array} \right]
 =
 M_0 
  \left[ \begin{array}{ccc}
   r_0 \\
   \theta_0
 \end{array} \right] \ .
\label{eq10}
\end{equation}
When the incident ray propagates parallel to the optical axis, $\theta_0 = 0$, as we shall assume here on, the exiting ray travels with an angle
\begin{equation}
 \theta = C_0 r_0
\label{eq10b}
\end{equation} 
in the output plane (and any other transverse plane of the focal region).
This expression should be interpreted as the paraxial approximation of Eq.~(\ref{eq02}).
Consequently, the focal distance is exclusively given by one element of the matrix, 
\begin{equation}
 f = - \frac{1}{C_0} \ .
\label{eq11}
\end{equation}
Negative values of $r$ and $\theta$ should be interpreted as an inversion of $180$~degrees with respect to the optical axis, accounted by an increment of $\pi$~rad in the azimuth angle.

Since the focal distance is dispersive by nature, $f(\omega)$, and thus the matrix element $C_0(\omega)$, a ray emerges from the optical system with an angle given by Eq.~(\ref{eq10b}) that varies for different frequencies.
If we consider a marginal ray propagating with a height $r_0 = R$, which denotes the maximum radial extent of the diffracting aperture, the emerging ray has a maximum angle $\alpha$, that is simply the (paraxial) numerical aperture of the focusing wave (in free space).
From Eq.~(\ref{eq10b}) we may obtain the dependence of the numerical aperture upon frequency,
\begin{equation}
 \frac{\alpha(\omega)}{\alpha_0} = \frac{C_0(\omega)}{C_0(\omega_0)} \ ,
\label{eq14}
\end{equation}
where $\alpha_0 = \alpha(\omega_0)$.

From Eq.~(\ref{eq10}) we have that the incident ray propagating parallel to the optical axis ($\theta_0 = 0$) emerges at the output plane with a height $r = A_0 r_0$.
In the future we consider the output plane corresponds to the back focal plane for a reference frequency $\omega_0$, and therefore $r(\omega_0) = 0$ and $A_0 (\omega_0) = 0$.
Longitudinal chromatic aberration makes that the position of the focal point varies with frequency, what implies that $r(\omega)$, and more importantly $A_0 (\omega)$, may differ from zero for a frequency other than $\omega_0$.
Therefore, we should consider the propagation from the output plane along a distance $z_0$ in order to meet the focus.
To find the value of $z_0$ we consider the matrix $M_0$ and a free-space propagation matrix accounted in the complete transfer matrix $M$ as
\begin{equation}
 \left[ \begin{array}{ccc}
   A & B \\
   C & D
 \end{array} \right]
 =
 \left[ \begin{array}{ccc}
   1 & z_0 \\
   0 & 1
 \end{array} \right]
 \left[ \begin{array}{ccc}
   A_0 & B_0 \\
   C_0 & D_0
 \end{array} \right] \ .
\label{eq12}
\end{equation}
In the back focal plane for a given frequency $\omega$, the value of the element $A(\omega)$ vanishes.
The solution of the linear equation $A = 0$ is 
\begin{equation}
 z_0 = - \frac{A_0}{C_0} \ .
\label{eq13}
\end{equation}
Thus, the longitudinal dispersion is determined from the elements $A_0$ and $C_0$ of the transfer matrix.
In conclusion, from Eqs.~(\ref{eq11}) and (\ref{eq13}) we infer that the spatial (and angular) dispersion of the focused beam is exclusively characterized by two elements of the $2 \times 2$-matrix $M_0$, $A_0$ and $C_0$.

In a general problem explored with the ray matrix method, we evaluate the four elements of the matrix $M_0$.
However, the determinant gives unity, 
\begin{equation}
 A_0 D_0 - B_0 C_0 = 1 \ ,
\label{eq14b}
\end{equation}
thus reducing the degrees of freedom.
Previously we have deduced that we only two elements of the matrix $M_0$, concretely $A_0$ and $C_0$, are necessary to account for the dispersive properties of the focused wave.
Consequently, the Debye representation derived in the previous section should impose a further constraint.
In this sense, note that the scalar diffraction integrals of Eqs.~(\ref{eq03}) and (\ref{eq07}) give an exact solution of the focused wavefield for telecentric lens systems \cite{Zapata00}. 
In this case, the aperturing screen is placed in the front focal plane, what represents the input plane.
Since $-D_0/C_0$ evaluates the distance from the input plane to the front focal plane, a telecentric optical system satisfies 
\begin{equation}
 D_0 = 0 \ . 
\end{equation}
Therefore, such a requirement additionally assures the exact validity of the Debye representation.
Unfortunately, wave dispersion frustrates that $D_0$ vanishes for a spectral range and, at least, we may impose $D_0$ to be sufficiently small, for example, to vanish for the reference frequency $\omega_0$.

Alternatively, the Debye approach is commonly examined in terms of the Fresnel number, $N = R_0^2 / \lambda |F_0|$ \cite{Collet80}.
In our case, $R_0$ denotes the radius of the exit pupil plane, which is conjugate of the input plane.
Also, $F_0$ represents the distance from the back focal plane to the exit pupil plane,
\begin{equation}
 F_0 = -\frac{B_0}{D_0} \ .
\label{eq14bb}
\end{equation}
Finally, the Fresnel number of the focusing geometry is written as
\begin{equation}
 N = \frac{(R/D_0)^2}{\lambda |F_0|} = \frac{R^2}{\lambda |B_0 D_0|} \ ,
 \label{eq14bc}
\end{equation}
where $\lambda = 2 \pi c / \omega$ is the wavelength.
In the previous equation we have used that the magnification corresponding to the input (pupil) plane is $D_0^{-1}$.
In telecentric optical systems, the value of $F_0$ is infinity and, since $D_0 = 0$, the Fresnel number also tends to infinity.
In the next section we will show that the Debye representation accurately reproduces the focused field when the Fresnel number reaches values much higher than unity, $N \gg 1$.

Finally, let us analyze the situation where the longitudinal dispersion, given in terms of the axial parameter $z_0$, is sufficiently small.
By using Eq.~(\ref{eq14b}) we may rewrite
\begin{equation}
 z_0 = \frac{A_0 B_0}{1 - A_0 D_0} \approx A_0 B_0 \ .
\label{eq14c}
\end{equation}
We have assumed that $A_0 (\omega_0) = 0$ and $D_0$ is small in the neighbourhood of $\omega_0$.
Under these circumstances, $z_0$ may be expanded into a power series of the term $A_0 D_0$.
In the lowest order we find the approximation $z_0 = A_0 B_0$, which is valid when
\begin{equation}
 |A_0 D_0| \ll 1 \ ,
\label{eq14d}
\end{equation}
that is, for a given frequency $\omega$ sufficiently close to the reference frequency $\omega_0$.
Eq.~(\ref{eq14c}) may be compared with Eq.~(\ref{eq13}), what gives the relation
\begin{equation}
 C_0 = - \frac{1}{B_0} \ ,
\label{eq14e}
\end{equation}
which is consistent with the approximation $|A_0 D_0| \ll 1$ together with the constraint of unitary determinant of $M_0$ given in Eq.~(\ref{eq14b}).

\section{Debye approximation of the Collins diffraction integral}

\subsection{Matrix-based formalism and region of validity of the Debye representation}

In terms of $ABCD$ matrices, the wavefield emerging from a diffraction-limited optical system may be determined from the Fresnel-Kirchhoff diffraction integral in the form of the Collins formula \cite{Siegman86}, 
\begin{eqnarray}
 U_\omega(\boldsymbol{r_\bot},z)=\frac{\omega}{i 2 \pi c} \frac{1}{B} 
  \exp{\left( i \frac{\omega}{2 c} \frac{D}{B} r^2 \right)} \nonumber \\
  \int \int_\infty^\infty U_\omega(\boldsymbol{r}_0)
  \exp{\left( i \frac{\omega}{2 c} \frac{A}{B} r_0^2 \right)}
  \exp{\left( -i \frac{\omega}{c} \frac{1}{B} \boldsymbol{r_\bot} \boldsymbol{r}_0 \right)}
  \mathrm{d}^2 \boldsymbol{r}_0 \ ,
\label{eq15}
\end{eqnarray}
where 
\begin{equation}
 \left[ \begin{array}{ccc}
   A & B \\
   C & D
 \end{array} \right]
 =
 \left[ \begin{array}{ccc}
   1 & z \\
   0 & 1
 \end{array} \right]
 \left[ \begin{array}{ccc}
   A_0 & B_0 \\
   C_0 & D_0
 \end{array} \right] \ .
\label{eq16}
\end{equation}
In Eq.~(\ref{eq15}), $U_\omega(\boldsymbol{r}_0)$ is the amplitude distribution in the input plane, where the aperture is placed, which accounts for the (occasionally dispersive) complex transmittance of the diffracting element and the spectral strength of the source.
Also, we have omitted a term $\exp{[i \omega (L + z) / c]}$, where $L$ is the axial distance from the input plane to the output plane, the latter being the back focal plane for a reference frequency $\omega_0$ (usually the carrier frequency of pulsed beams or the mean frequency of the power spectrum for incoherent white light).
Moreover, the spatial coordinate $z$ represents an axial distance from the output plane, and $\boldsymbol{r_\bot}=(r \cos \phi,r \sin \phi)$ indicates the transverse spatial coordinates. 
Finally, the matrix $M_0$ is evaluated from the input plane to the output plane, which elements are, in general, dispersive and therefore depending on frequency.
The particular selection of the output plane guaranties that $A_0(\omega_0) = 0$.

Eqs.~(\ref{eq07}) and (\ref{eq15}) allow us to evaluate the 3-D amplitude distribution in the focal region.
However, comparison of the paraxial diffraction integral within the Debye approximation and the Collins formula manifests some differences.
First we address our attention to the dependence upon the axial spatial coordinate of the phase terms in the diffraction integral.
In the Debye representation, the argument of the phase terms varies linearly with $z$, in opposition to the more complicated dependence in the Collins integral.
An approach may be performed after a first-order series expansion of the argument about the point $z = 0$.
Concretely, we may write
\begin{equation}
 \frac{A}{B} = \frac{A_0}{B_0} - \frac{z}{B_0 (B_0 + z D_0)} \approx 
  \frac{A_0}{B_0} - \frac{z}{B_0^2} \ ,
 \label{eq17}
\end{equation}
where $A_0 D_0 - B_0 C_0 = 1$ of Eq.~(\ref{eq14b}) has been used.
The physical interpretation of this approximation lies on the assumption that the focal field is mostly concentrated in the vicinity of focus, satisfying
\begin{equation}
 |z| \ll \left|\frac{B_0}{D_0} \right| \ ,
 \label{eq18}
\end{equation}
and then we may write $B = B_0 + z D_0 \approx B_0$.
Moreover, in terms of the axial parameter $F_0 = - B_0 / D_0$ given in Eq.~(\ref{eq14bb}), which represents the distance from the back focal plane to the image of the input plane (exit pupil plane), the previous inequality may be simplified as $|z| \ll |F_0|$.
Consequently, Eq.~(\ref{eq18}) involves that the focal volume is far from the exit pupil plane.

We identify a second difference in the presence of a quadratic (on $r$) phase factor in Eq.~(\ref{eq15}), external to the diffraction integral, which is not found in Eq.~(\ref{eq07}).
However, this term may be neglected under the assumption that the evaluation of the focal field is performed in a region where
$(\omega / 2 c) |D / B| r^2 \ll \pi$, that is,
\begin{equation}
 \frac{r^2}{\lambda |F_0|} \ll 1 \ .
 \label{eq19}
\end{equation} 
The left side of this inequality has the form of a Fresnel number, and next it will be analyzed in detail.
In resume, a full analogy of the diffraction integrals given from the Fresnel-Kirchhoff formulation and the Debye representation is found if the analysis of the focal field is accomplished in the restricted region given by Eqs.~(\ref{eq18}) and (\ref{eq19}). 

Let us write the Debye diffraction integral in the $ABCD$ matrix representation,
\begin{eqnarray}
 U_\omega(\boldsymbol{r_\bot},z)=\frac{\omega}{i 2 \pi c} \frac{1}{B_0} 
  \int \int_{-\infty}^\infty U_\omega(\boldsymbol{r}_0)
  \exp{\left( i \frac{\omega}{2 c} \frac{A_0}{B_0} r_0^2 \right)} \nonumber \\
  \exp{\left( -i \frac{\omega}{2 c} \frac{z}{B_0^2} r_0^2 \right)} 
  \exp{\left( -i \frac{\omega}{c} \frac{1}{B_0} \boldsymbol{r_\bot} \boldsymbol{r}_0 \right)}
  \mathrm{d}^2 \boldsymbol{r}_0 \ .
 \label{eq21}
\end{eqnarray}
In comparison with Eq.~(\ref{eq07}) we find that the focal distance is written as $f = B_0$.
From geometrical considerations we have obtained the same result written in Eq.~(\ref{eq11}), combined with the approximated equation~(\ref{eq14e}).
From the comparison we also have that $z_0 / f^2 = A_0 / B_0$, that is, $z_0 = A_0 B_0$. 
Again, this result has been previously obtained in Eq.~(\ref{eq14c}) under the assumption that $|A_0 D_0| \ll 1$.
Alternatively, this inequality may be written as
\begin{equation}
 |z_0 / F_0| \ll 1 \ ,
 \label{eq22}
\end{equation}
which is simply the condition given in Eq.~(\ref{eq18}) for the specific value of the axial coordinate $z = z_0$.
In consequence, the focus position observed for a given frequency, $F(\omega)$, is to be far from the exit pupil plane, that is, in the near-field region of the focused wave.
In the case of highly-dispersive focusing elements, this condition may bring serious restrictions on the spectral bandwidth of the incident beam in order to assure the validity of the Debye representation.

We point out that when $D_0 = 0$, the Collins formula given in Eq.~(\ref{eq15}) reduces to the matrix-based diffraction integral in the Debye representation of Eq.~(\ref{eq21}).
In this case $F_0$ tends to infinity, that is, the exit pupil is located infinitely far away from the focal region.
As a consequence, the validity of the Debye formulation is expanded into the whole space.
In this singular case one cannot speak of ``Debye approximation''.

Importantly, we may conclude that the Debye approximation imposes bounds on the spatial coordinates where the focal wavefield is evaluated, as seen in Eqs.~(\ref{eq18}) and (\ref{eq19}), giving an accurate estimation under a limited longitudinal dispersion shown in Eq.~(\ref{eq22}).

\subsection{Spatial symmetries}

A relevant issue associated with apertured spherical waves correctly described in the Debye representation is the inherent symmetry about the focal point.
From Eq.~(\ref{eq21}) we find that if the wave function in the input plane is real, $U_\omega (\mathbf{r}_0) \in \Re$, then the focal field is symmetric with respect to the focal point, 
\begin{equation}
 U_\omega(\boldsymbol{r_\bot},z + A_0 B_0) = 
     - U^*_\omega(- \boldsymbol{r_\bot},-z + A_0 B_0) \ ,
\label{eq31}
\end{equation}
where $*$ denotes a complex conjugate.
Note that we have considered the focal point is found at a distance $z_0 = A_0 B_0$.
A real wave function is encountered, for example, when a pulsed unchirped Gaussian beam propagates through a diffraction-limited optical system with a clear circular aperture.
In particular, Wolf \textit{et al.} \cite{Collet80,Wolf81} considered the spatial symmetries of the intensity in the focal region of polychromatic spherical waves.
They concluded that when the diffracting screen is purely absorbing, the focal intensity is centro-symmetric about the focus, where the maximum intensity is observed.

For the sake of simplicity, we follow our discussion under the hypothesis that the wavefield in the input plane may be factorized in the form $U_\omega(\boldsymbol{r}_0) = S_0(\omega) T(r_0)$.
The first function represents the spectral strength of the incident radiation, which depends exclusively on the frequency.
The second is an azimuthally-symmetric function of the radial spatial coordinate, which accounts for the aperture transmittance.
In this diffraction problem we assume that the amplitude transmittance of the diffracting aperture is frequency independent, what holds for purely-absorbing screens.
The Collins diffraction formula in the Debye approximation is finally written as 
\begin{eqnarray}
 U_\omega(\mathbf{r})=\frac{\omega}{i c} \frac{S_0}{B_0} 
  \int_0^R T(r_0)
  \exp{\left( i \frac{\omega}{2 c} \frac{A_0}{B_0} r_0^2 \right)} \nonumber \\
  \exp{\left( -i \frac{\omega}{2 c} \frac{z}{B_0^2} r_0^2 \right)} 
  \mathrm{J}_0 \left( \frac{\omega}{c} \frac{1}{B_0} r r_0 \right)
  r_0 \mathrm{d} r_0 \ ,
 \label{eq23}
\end{eqnarray}
where $R$ is the radius of the aperture.

\subsection{The role of the Fresnel number}

Previously we have demonstrated that the diffraction integral in the Debye representation may be systematically employed with a unique spatial limitation (apart from the longitudinal dispersion), conceiving a region of validity shown in Eqs.~(\ref{eq18}) and (\ref{eq19}). 
In principle, we lack the guaranty that the region of interest, that is, the focal volume, belongs to the region of validity of the Debye representation. 
Here we analyze the conditions an spherical wavefield, that emerges from an apertured optical system of transfer matrix $M_0$, should satisfy in order to enclose the focal volume in the required bounded region.

For simplicity, let us analyze the case of a diffraction-limited optical arrangement with a circular clear aperture of radius $R$, i.e., $T(r_0) = 1$ \cite{Sheppard88Martinez99}.
Also, in this examination we neglect the longitudinal dispersion, and then we restrict our analysis to the case where $A_0 = 0$ for a given frequency.
The field distribution along the optical axis is calculated by means of Eq.~(\ref{eq23}), which particularized to $r = 0$ gives
\begin{equation}
 U_\omega(z)= \frac{S_0 B_0}{z} \left\{ 
 \exp{\left( -i \frac{\omega}{B_0^2} \frac{z}{2 c} R^2 \right)} - 1
 \right\} \ .
 \label{eq23b}
\end{equation}
The on-axis intensity $|U_\omega(z)|^2$ is maximum at the origin, and the first zeros are found at the points of axial coordinates
\begin{equation}
 z_1 = \frac{4 \pi c}{R^2} \frac{B_0^2(\omega)}{\omega} \ ,
 \label{eq23c}
\end{equation}
and $- z_1$.
The highest values of the on-axis intensity are found at points comprised between these two zeros, and thus we may consider the region of interest is $|z| \le z_1$.
Validity of the Debye representation in this region implies that the inequality of Eq.~(\ref{eq18}) is satisfied for $|z| \le z_1$, what means that $z_1 \ll |B_0 / D_0|$.
Equivalently we may write
\begin{equation}
 2 \ll \frac{R^2}{\lambda |B_0 D_0|} \ ,
 \label{eq23d}
\end{equation}
that is, $2 \ll N$, where $N$ is the Fresnel number given in Eq.~(\ref{eq14bc}).
In accordance we conclude that a Fresnel number much higher than unity guaranties that the focal field is accurately described with the Debye representation along the optical axis and neighbouring points.

On regards the off-axis points of the focal volume, let us investigate the wavefield in the transverse focal plane, $z = 0$.
From Eq.~(\ref{eq23}) we finally have
\begin{equation}
 U_\omega(r)= \frac{S_0 R}{i r} \mathrm{J_1} \left( 
 \frac{\omega}{B_0} \frac{r}{c} R 
 \right) \ .
 \label{eq23e}
\end{equation}
where $\mathrm{J_1}$ is a Bessel function of the first kind and order 1.
The field distribution given in Eq.~(\ref{eq23e}) is simply the so-called Airy disk.
Again, the maximum of intensity occurs at the origin, and the first zero of intensity is found at
\begin{equation}
 r_1 = 1.22 \pi \frac{c}{R} \frac{|B_0|}{\omega} \ .
 \label{eq23f}
\end{equation}
Note also that $r_1$ gives the limit of resolution of a diffraction-limited imaging system in agreement with the Rayleigh criterion.
Again, the Debye representation is valid if Eq.~(\ref{eq19}) is fully satisfied for points $r \le r_1$.
Thus, we should impose that $r_1^2 \ll \lambda |B_0 / D_0|$, what may be written as 
$0.37 \ll N$.
As a result, high values of the Fresnel number allow the focal field to be confined in the region of validity of the Debye representation, satisfying simultaneously the inequalities of Eqs.~(\ref{eq18}) and (\ref{eq19}).

\section{Examples}

\subsection{Apertured thin lens}

\begin{figure}
\includegraphics[width=8cm]{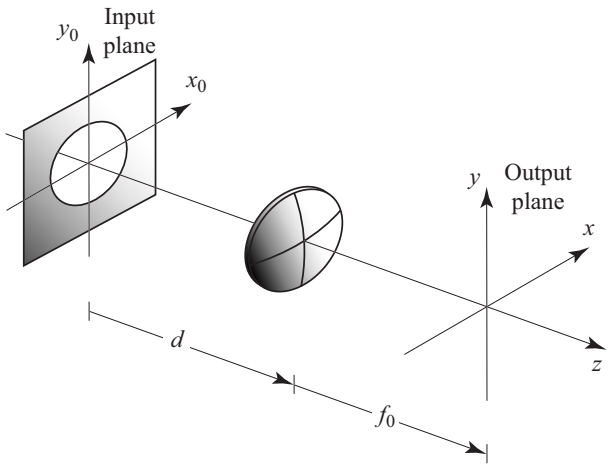}
\caption{Diagram illustrating an apertured thin lens.
 The diffracting aperture is placed at a distance $d$ in front of the lens.
 The lens focal length is $f_0$ for the reference frequency $\omega_0$.}
\label{fig02}
\end{figure}

Let us first consider a positive thin lens of focal distance $f(\omega)$.
The term $f_0$ is assigned to the value of the focal distance for the reference frequency, i.e., $f_0 = f(\omega_0)$.
A diffracting screen is placed in front of the lens at a distance $d$, as seen in Fig.~\ref{fig02}.
A negative value of $d$ may consider an aperture located at the back of the lens.
As previously established, the input plane is set in the place of the diffracting aperture, whereas the output plane corresponds to the back focal plane for $\omega_0$.
In this case, the transfer matrix $M_0$ has the following elements:
\begin{subequations}
 \label{eq24}
 \begin{eqnarray}
  A_0(\omega) &=& 1 - \frac{f_0}{f(\omega)} \ , \label{eq24a} \\
  B_0(\omega) &=& f_0 + d - \frac{d f_0}{f(\omega)} \ , \label{eq24b} \\
  D_0(\omega) &=& 1 - \frac{d}{f(\omega)} \ , \label{eq24c}
 \end{eqnarray}
\end{subequations}
and, obviously, $C_0(\omega) = - f^{-1}(\omega)$.
Supposing that the focal distance is independent upon frequency, what occurs for achromatic objectives, the term $A_0$ takes values identically zero.
However, $A_0$ may reach significant values when the focusing lens is highly dispersive, thus producing a notable longitudinal dispersion.
Also, when the diffracting aperture is at the front focal plane for a given frequency $\omega$, being $d = f(\omega)$, the focusing system is telecentric and, correspondingly, $D_0 = 0$.

\begin{figure}
\includegraphics[width=8cm]{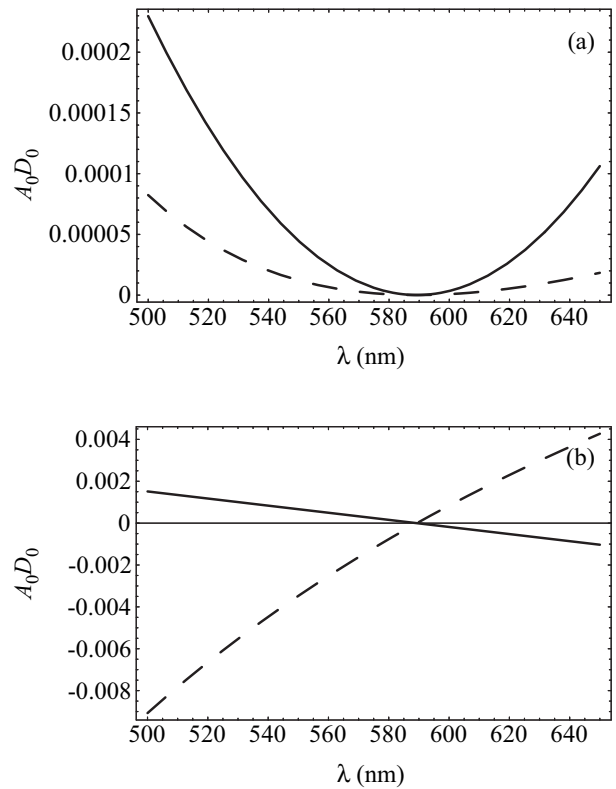}
\caption{Wavelength dependence of the product $A_0 D_0$ in a central band of the visible spectrum.
 The elements of the transfer matrix are evaluated for a Fresnel lens (solid line) and a BK7 glass lens (dashed line) with the same focal length $f_0 = 10$~cm for a wavelength $\lambda_0 = 589.3$~nm.
 The aperture is placed at: (a) $d = f_0$, and (b) $d = 0$.
 The values for the Fresnel lens have been multiplied by a factor of $10^{-2}$.}
\label{fig03}
\end{figure}

\begin{table}
 \caption{Sellmeier coefficients for a common borosilicate crown (BK7) glass \cite{Sellmeier}.}
 \label{tab:01}
 \begin{ruledtabular}
  \begin{tabular}{ll}
   Coefficient & Value\\
   B1 & 1.03961212\\
   B2 & 2.31792344 $10^{-1}$ \\
   B3 & 1.01046945 \\
   C1 & 6.00069867 $10^{-3}$~$\mu$m$^2$ \\
   C2 & 2.00179144 $10^{-2}$~$\mu$m$^2$ \\
   C3 & 1.03560653 $10^2$~$\mu$m$^2$ \\
   \end{tabular}
  \end{ruledtabular}
\end{table}

Two sort of thin lenses are analyzed in this section: kinoform-type diffractive lenses and refractive thin lenses.
Diffractive singlets are zone plates which may achieve a high-efficiency performance with phase-only multilevels \cite{Levi01}.
Recently, some authors have addressed special attention to investigate the spatio-temporal response of broadband ultra-short pulses focused with zone plates \cite{Pearce02,Ashman03,Zapata06c}.
Here we are giving a simplified formalism to evaluate the focal field in the frame of the Debye representation.
The focal length of a Fresnel lens may be modelized as \cite{Moreno97}
\begin{equation}
 \frac{f_0}{f(\omega)} = \frac{\omega_0}{\omega} \ .
\label{eq26}
\end{equation}
Secondly, quartz and glass thin lenses are considered.
Plano convex lenses are the most basic optical elements.
They have positive focal lengths and close to the optimum shape for use as focusing lenses for collimated beams.
If $n(\omega)$ denotes the refractive index of the transparent material the lens is made of, the focal length of the refractive singlet may be expressed as
\begin{equation}
 \frac{f_0}{f(\omega)} = \frac{n(\omega) - 1}{n_0 - 1} \ ,
\label{eq27}
\end{equation}
being $n_0 = n(\omega_0)$. 
Bi-convex and other spherical lenses may be treated similarly.
Instead of using tabulated data of the refractive index $n(\omega)$ of a given medium, we may employ the Sellmeier equation, which is an empirical relationship between refractive index $n$ and wavelength $\lambda$.
The usual form of the equation for glasses is:
\begin{equation}
 n^2(\lambda) = 1 + \frac{B_1 \lambda^2}{\lambda^2 - C_1} + 
  \frac{B_2 \lambda^2}{\lambda^2 - C_2} + \frac{B_3 \lambda^2}{\lambda^2 - C_3} \ ,
\label{eq28}
\end{equation}
where $B_i$ and $C_i$, for $i=1,2,3$, are experimentally determined Sellmeier coefficients.
Note that this $\lambda = 2 \pi c / \omega$ is the vacuum wavelength.
As an example, here we consider a common borosilicate crown glass known as BK7, and the Sellmeier coefficients are shown in Table~\ref{tab:01}.


In Fig.~\ref{fig03} we investigate the longitudinal chromatic aberration in relation to proximity to the exit pupil plane.
This point is important since the Debye approximation is restricted to a limited axial dispersion expressed as $|A_0 D_0| \ll 1$, as given in Eq.~(\ref{eq22}).
In the figure we use the reference frequency $\omega_0$ associated to the average wavelength $\lambda_0 = 589.3$~nm of the Sodium doublet.
In general, the highly-dispersive character of the Fresnel lens severely restricts the spectral range we may use the Debye representation of the focal field.
A relevant property is that when $d = f_0$, i.e. the system is telecentric for the radiation frequency $\omega_0$, the term $A_0 D_0$ is stationary.
In other words, $A_0 D_0 = 0$ for the reference frequency $\omega_0$ (since $A_0(\omega_0) = 0$) and
\begin{equation}
 \frac{\partial (A_0 D_0)}{\partial \omega}  = 0 \ ,
\label{eq26b}
\end{equation}
for $\omega_0$.
Consequently, the values of $A_0 D_0$ remain significantly low in the spectral neighbourhood around $\omega_0$, as seen in Fig.~\ref{fig03}(a).
However, when the aperture is at the lens plane ($d = 0$), $\omega_0$ is not a stationary point of $A_0 D_0$, whose values may increase up to one order of magnitude, as shown in Fig.~\ref{fig03}(b).
The analysis of a BK7 glass thin lens has also been included in the figure.
Comparatively, the refractive lens gives values of $|A_0 D_0|$ considerably lower than those provided by a Fresnel lens.
In this case, application of the Debye approximation is less stringent in terms of bandwidth, even considering the case of apertures located at the lens plane.

\begin{figure}
\includegraphics[width=8cm]{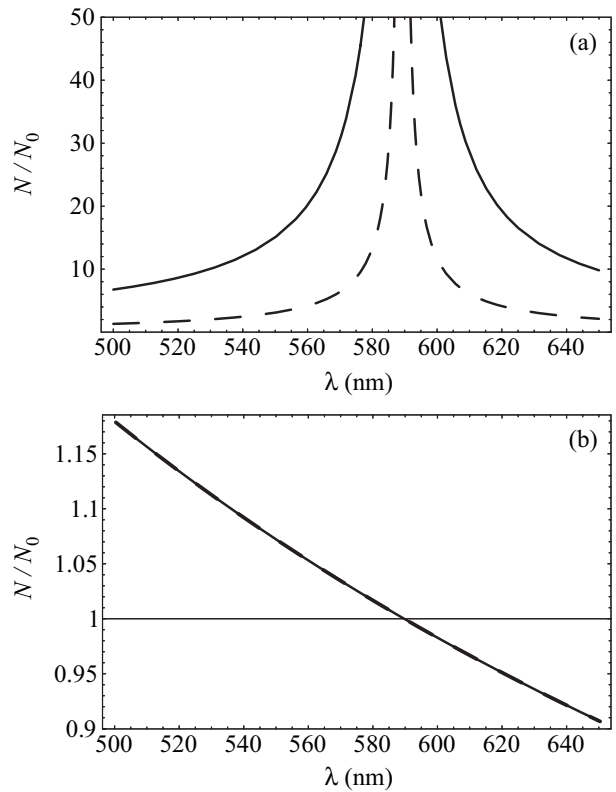}
\caption{Spectral dependence of the reduced Fresnel number $N / N_0$ when the diffracting screen is placed at (a) $d = f_0$, and (b) $d = 0$.
 The elements of the transfer matrix are evaluated for a Fresnel lens (solid line) and a BK7 glass lens (dashed line).
 The values for the Fresnel lens have been multiplied by $10^{-2}$ in subfigure (a).}
\label{fig04}
\end{figure}

Validity of the Debye approximation is also restricted to Fresnel numbers much higher than 1, what is under study in the following example.
Consider a circular clear aperture of diameter $2 R = 1$~cm and a thin lens of focal length $f_0 = 10$~cm for the reference wavelength $\lambda_0 = 589.3$~nm.
If the diffracting aperture is placed at the lens plane, the Fresnel number for the reference frequency
\begin{equation}
 N_0 = \frac{R^2}{\lambda_0 f_0} \ .
\label{eq29}
\end{equation}
In the present numerical example $N_0 = 424$.
For other frequencies and positions of the aperture, the Fresnel number $N(\omega)$ varies according to Eq.~(\ref{eq14bc}).
Let us examine a reduced Fresnel number, normalized to the Fresnel number $N_0$, written as
\begin{equation}
 \frac{N}{N_0} = \frac{\omega}{\omega_0} \frac{f_0 }{|B_0(\omega)|} \frac{1}{|D_0(\omega)|} \ .
\label{eq30}
\end{equation}
When the aperture is placed at the lens plane, $d = 0$, the terms $B_0 = f_0$ and $D_0 = 1$ are spectrally invariant (non-dispersive), and the reduced Fresnel number is finally expressed as $N/N_0 = \omega/\omega_0$. 
This result is valid for any thin (Fresnel and dispersive glass) lens, as seen in Fig.~\ref{fig04}(b).
In the case the focusing system is telecentric (for $\omega_0$), where $d = f_0$, the value of $D_0(\omega_0)$ vanishes and, consequently, the Fresnel number diverges for the reference frequency.
As shown in Fig.~\ref{fig04}(a), the Fresnel number for neighboring frequencies reaches extremely high values.
Comparatively, the Fresnel number associated with the Fresnel lens are several orders of magnitude higher that those of a BK7 glass lens.
In general, we may conclude that the requirement $N \gg 1$ is fully satisfied within the visible spectral band.

\begin{figure}
\includegraphics[width=8cm]{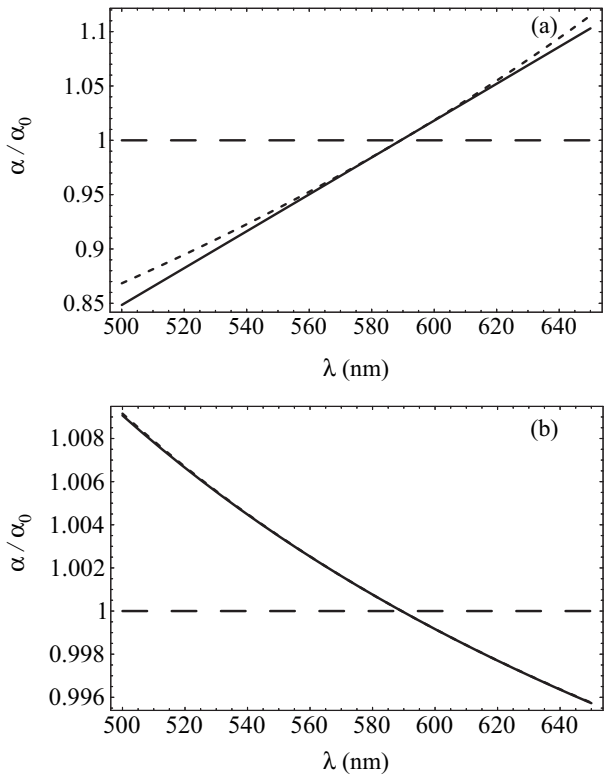}
\caption{Wavelength dependence of the relative numerical aperture $\alpha / \alpha_0$ for: (a) a Fresnel lens and (b) a BK7 glass lens.
 The curves are evaluated by means of Eq.~(\ref{eq14}) (solid line), and Eq.~(\ref{eq30b}) for $d = 0$ (dashed line) and $d = f_0$ (dotted line).}
\label{fig05}
\end{figure}

The angular dispersion of the focal waves produced by a thin lens may be examined with the matrix element $C_0(\omega)$ and Eq.~(\ref{eq14}).
For instance, a diffractive lens has high-dispersive numerical aperture given by
\begin{equation}
 \frac{\alpha(\omega)}{\alpha_0} = \frac{\omega_0}{\omega} \ ,
\end{equation}
irrespective of aperture position.
Alternatively we may use that $B_0 C_0 \approx -1$ in the Debye approximation, and thus the spectral dependence of the numerical aperture may be expressed as
\begin{equation}
 \frac{\alpha(\omega)}{\alpha_0} = \frac{B_0(\omega_0)}{B_0(\omega)} \ .
 \label{eq30b}
\end{equation}
In Fig.~\ref{fig05} we plot the relative numerical aperture for a diffractive lens (subfigure a) and a BK7 glass lens (subfigure b).
The exact expression given in Eq.~(\ref{eq14}) shows that the relative numerical aperture, and thus the angular dispersion, is independent of the variable $d$, i.e., the aperture place.
However, Eq.~(\ref{eq30b}) reveals a small dependence on $d$, mainly in the case of the zone plate.
This point is important since we deduce from Eq.~(\ref{eq30b}) that, in the case $d = 0$, the numerical aperture is nondispersive, i.e., $\alpha(\omega) = \alpha_0$. 
Strictly speaking, this result is false though the angular dispersion is relatively small and may be ignored.
On the contrary, the approximation given in Eq.~(\ref{eq30b}) is extremely accurate when $d = f_0$.
We may conclude that deviations of Eqs.~(\ref{eq14}) and (\ref{eq30b}) intrinsically elucidate the degree of accuracy of the Debye approximation.

\begin{figure*}
\includegraphics[width=17cm]{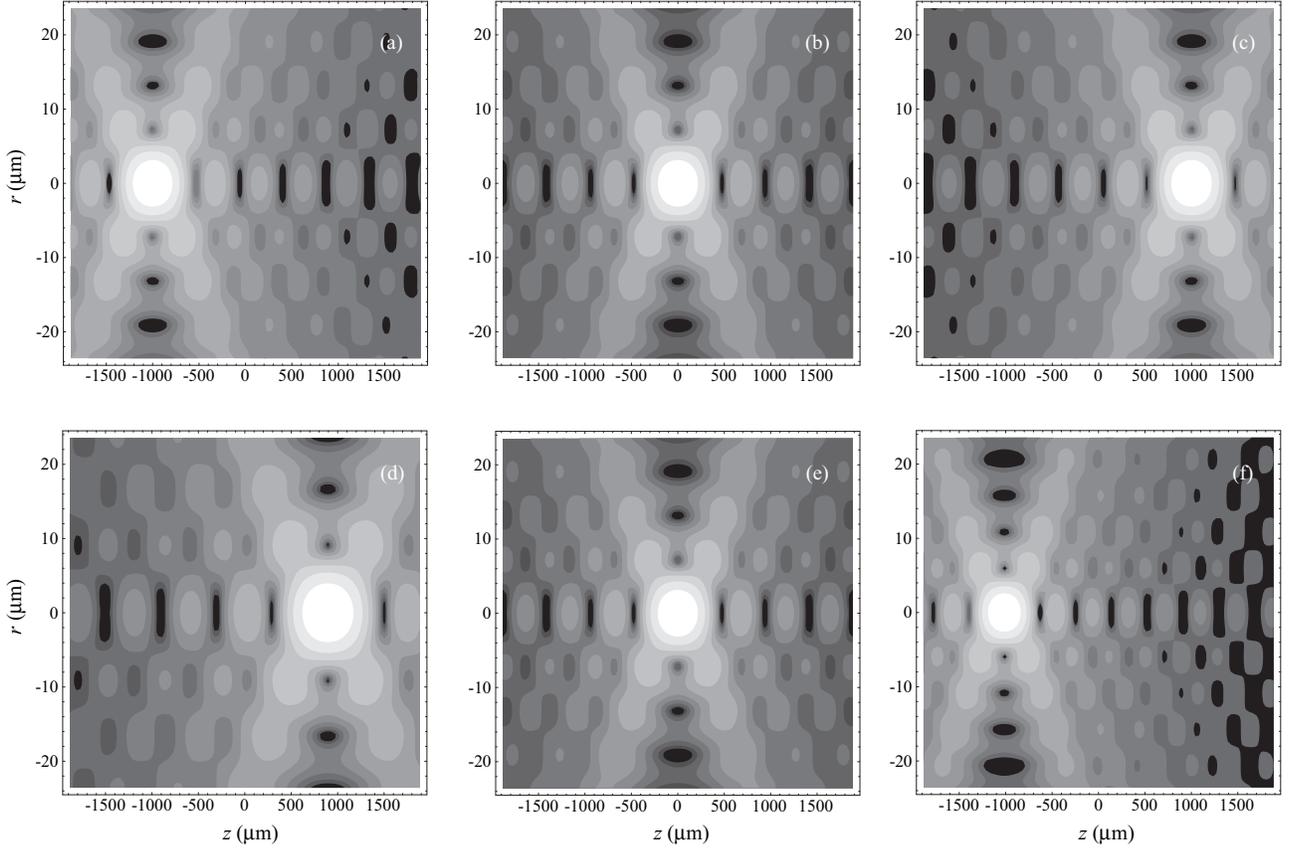}
\caption{Contour plot of the field strength in log scale (natural logarithm of $|U_\omega (\mathbf{r})|^2$) in the focal region of (a-c) a Fresnel lens and (d-f) a BK7 glass lens, both of focal length $f_0 = 10$~cm and with an aperture placed at $d = f_0$.
 We consider a frequency (a, d) $(1 - \Delta) \omega_0$, (b, e) $\omega_0$, and (c, f) $(1 + \Delta) \omega_0$.
 In this case $\Delta = 0.01$ for the Fresnel lens, and $\Delta = 0.2$ for the glass lens.
The contour interval (difference between successive contour lines) is 0.75.}
\label{fig06}
\end{figure*}


In Fig.~\ref{fig06} we show some contour plots of the three-dimensional intensity distribution of an apertured Fresnel thin lens and a BK7 glass lens of focal distance $f_0 = 10$~cm for a wavelength $\lambda_0 = 589.3$~nm.
The field intensity is numerically evaluated from Eq.~(\ref{eq23}) for $T(r_0) = 1$ and different frequencies.
Also, we have selected a value of $S_0$ such that the intensity at the focus is unity, and the gray levels are plotted in a logarithmic scale of intensities to increase the contrast.
Again, the diffracting screen is a clear circular aperture of diameter $2 R = 1$~cm and placed at a distance $d = f_0$.
The predicted symmetries of the intensity around the geometrical focus are evident in the plots.
In general, we observe that the glass lens presents a higher resistance to dispersion-induced focal shifts.
Additionally, for increasing frequencies, the focal shift is driven toward the lens for the refracting singlet, in opposition to the case of the Fresnel lens.
Finally, an additional spatial dispersion effect is distinct in the subfigures of the glass lens.
In order to have comparable longitudinal dispersion with the Fresnel lens, the selected frequencies considerably differ, with 20 per cent of increment ($\Delta \omega / \omega_0 = 0.2$).
In this case, the size of the focal spot is notably higher for lower frequencies, what is clearly observed by computing the number of minima (black shades) along the optical axis and in the transverse focal plane.

\subsection{White-light focusing system with compensated spatial dispersion}

A number of designs of achromatic doublet lenses \cite{Fischer00b} and hybrid refractive-diffractive achromats \cite{Ibragimov95} have been proposed to get rid of longitudinal chromatic aberration inherent in singlet lenses.
Ray matrices of this sort of focusing setups have (at least approximately) a vanishing term $A_0(\omega) = 0$, a performance that can be achieved over a broad spectral band.
Commonly, this property is accompanied with spectral invariance of the focal length, meaning that both $B_0$ and $C_0$ are also invariant in the considered spectral range.
When used on-axis, an achromatic lens focuses a parallel input beam to a single point.
However, the off-axis performance is significantly worse than the on-axis performance since it is limited by the effects of diffraction. 
For instance, the limit of resolution $r_1$ given in Eq.~(\ref{eq23f}) and frequency are inversely proportional.

\begin{figure}
\includegraphics[width=8cm]{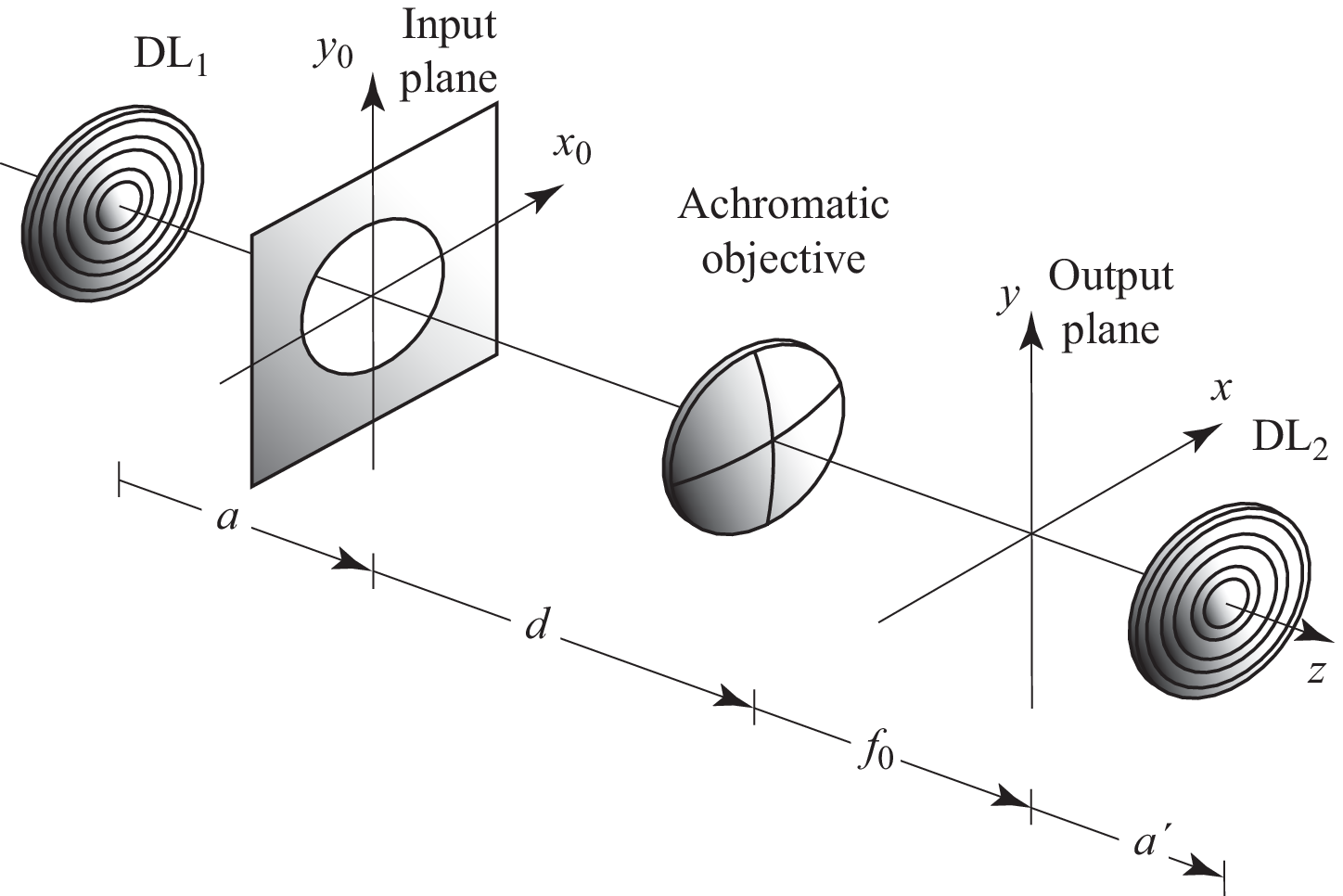}
\caption{Optical system with compensated spatial dispersion.
 The diffracting aperture is placed at a distance $d$ in front of the achromatic objective of focal length $f_0$.
 The appropriate insertion of a pair of kinoform-type diffractive lenses (DL$_1$ and DL$_2$) is able to strongly alter the angular dispersion of the focused beam without introducing longitudinal chromatic aberration.}
\label{fig08}
\end{figure}

In Ref. \cite{Lancis04} Lancis \textit{et al.} proposed an optical system composed of a nondispersive objective coupled with a pair of kinoform-type diffractive lenses to overcome longitudinal chromatic aberration and, additionally, the diffraction-induced spatial dispersion of the Fraunhofer pattern observed in the transverse focal plane.
A first diffractive lens (DL$_1$) of focal length $Z_1$ (for $\omega_0$) is placed in front of the achromat (of focal length $f_0$ for all frequencies), at a distance $s = a + d$, and the second diffractive lens (DL$_2$) of focal length $Z_2$ is located behind, at a distance $s' = f_0 + a'$ as depicted in Fig.~\ref{fig08}.
Under the conditions $s^{-1} + s'^{-1} = f_0^{-1}$ (lens formula) and $Z_1 Z_2 = -(s'/s)^2$, the emerging beam is focused at a distance $f_0$ from the achromatic objective, specifying the output plane.
In this case, $A_0(\omega) = 0$ in the spectral domain.
Additionally, inserting the diffracting aperture at a distance 
\begin{equation}
 a = \frac{Z_1}{2} 
\end{equation}
from the first diffracting lens, the rest of matrix elements are
\begin{subequations}
 \label{eq31}
 \begin{eqnarray}
  B_0(\omega) &=& f_0 \left(1 - \frac{\omega_0}{2 \omega} \right) \ , \label{eq31a} \\
  D_0(\omega) &=& 1 - \frac{d}{f_0} + \frac{Z_1 + 2 \left( d - f_0 \right)}{4 f_0} \frac{\omega_0}{\omega} \ , \label{eq31b}
 \end{eqnarray}
\end{subequations}
and, obviously, $C_0 = -B_0^{-1}$ (see Eq.~(\ref{eq14b})).
In Fig.~\ref{fig08} $a'$ is positive and, therefore, the focus is virtual.
However, inclusion of imaging lenses in the rear may generate a real focal plane.

\begin{figure}
\includegraphics[width=8cm]{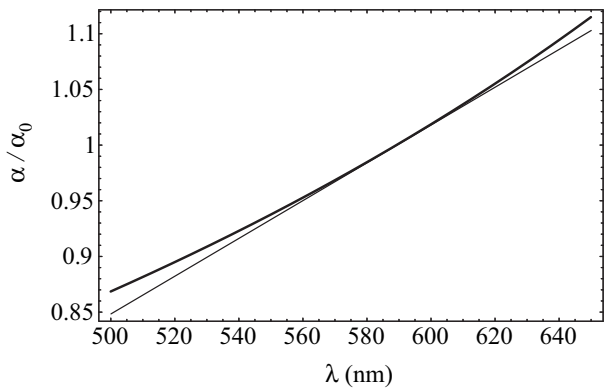}
\caption{Wavelength dependence of the relative numerical aperture for the system in Fig.~\ref{fig08}.
 The thin solid line corresponds to a system with perfectly-compensated spatial dispersion.
 In the latter case, the numerical aperture has a linear dependence with the wavelength, $\alpha / \alpha_0 = \lambda / \lambda_0$.
 In the plot we assume $\lambda_0 = 589.3$~nm.}
\label{fig12}
\end{figure}

Eq.~(\ref{eq31a}) indicates that the transfer matrix of the proposed system have a highly-dispersive element $B_0(\omega)$.
Fig.~\ref{fig12} depicts the spectral dependence of the term $B_0^{-1}$, that is, the angular dispersion of the focused waves as deduced from Eq.~(\ref{eq30b}).
We point out that Eq.~(\ref{eq30b}) evaluates the paraxial numerical aperture exactly since $A_0$ vanishes.
For frequencies sufficiently close to $\omega_0$ we find that 
\begin{equation}
 \frac{\alpha(\omega)}{\alpha_0} \approx \frac{\omega_0}{\omega} \ ,
\end{equation}
as occurring with a single zone plate.
In other words, the numerical aperture depends approximately linearly upon wavelength.
However, whereas a zone plate achieves this characteristic angular dispersion at the cost of longitudinal chromatic aberration, the proposed system has a frequency-independent focal plane.
As demonstrated in Ref.~\cite{Lancis04}, this fact is a direct consequence of imposing the reference frequency $\omega_0$ to be a stationary point for the term $\omega / B_0$, that is,
\begin{equation}
 \frac{\partial}{\partial \omega} \left( \frac{\omega}{B_0(\omega)} \right) = 0
\end{equation}
for $\omega_0$.
This allows a first-order achromatization of the Fraunhofer diffraction pattern.
Finally, when the diffracting screen is a clear aperture, the spectral components of the field (Airy disk) has a nearly invariant size and, according to Eq.~(\ref{eq23f}), a frequency-independent limit of resolution $r_1$.

\begin{figure}
\includegraphics[width=8cm]{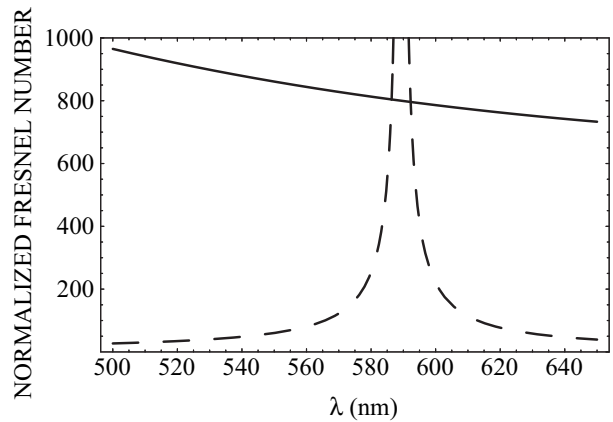}
\caption{Spectral dependence of the Fresnel number $N(\omega)$, normalized to the factor $N_0 (f_0/Z_1)$, when the diffracting screen is placed at $d = f_0$ (solid line), and $d = f_0 + Z_1 / 2$ (dashed line).
 For the sake of clarity, the values corresponding to the solid line have been multiplied by $10^{2}$.}
\label{fig11}
\end{figure}

The angular dispersion of this singular optical arrangement is independent of the parameter $d$, that is, the distance from the aperture to the achromatic lens may be arbitrarily changed.
However, the Fresnel number may vary substantially for different values of $d$.
The Fresnel number given in Eqs.~(\ref{eq29}) and (\ref{eq30}) is plotted versus wavelength for $d = f_0$ and $d = f_0 + Z_1 / 2$ in Fig.~\ref{fig11}.
In the first case (plotted in solid line), the optical system has a finite Fresnel number for the reference frequency $\omega_0$ and, therefore, locating the aperture at the front focal plane of the objective does not guarantee a telecentric lens design.
Contrarily, the second case (plotted in dashed line) accounts for the condition of telecentricity for $\omega_0$, for which the Fresnel number diverges.
This favours that the spatial constraints of the focal volume, associated with the Debye approximation, are fulfilled.

\begin{figure*}
\includegraphics[width=17cm]{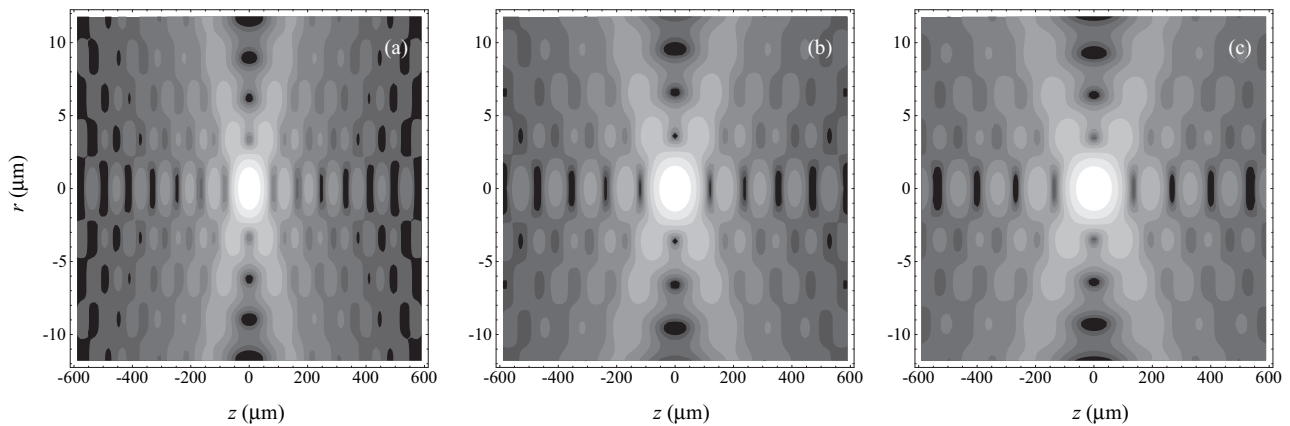}
\caption{Contour plot of the field strength  in the focal region of the optical system in Fig.~\ref{fig08}. 
 The contour intervals vary $0.75$ in a log scale to augment the contrast.
 The achromatic objective has a focal length $f = 10$~cm, and the aperture is placed at a distance $d = 10$~cm.
 Each plot corresponds to the frequency (a) $(1 - \Delta) \omega_0$, (b) $\omega_0$, and (c) $(1 + \Delta) \omega_0$, with $\Delta = 0.2$.}
\label{fig09}
\end{figure*}

Finally, Fig.~\ref{fig09} depicts the isophotes in the neighbourhood of the focal point for different frequencies.
In all cases, the Airy disk formed in the focal plane has a comparable magnitude.
However, an increase in frequency induces a broadening of the focal spot along the optical axis.
We point out that this behavior substantially differs from a conventional achromat lens, where transverse and axial widths of the central lobe have a linear dependence with the inverse of the frequency.


\section{Conclusions}

We have derived an ABCD matrix-based generalization of the Debye representation of focal fields.
This diffraction integral has been achieved using two alternative procedures.
In the first approach we substitute the focal length and the longitudinal dispersion, which are explicit parameters in the classical Debye integral, by equivalent expressions given in terms of the ray-matrix elements.
The second method consists in applying simple approximations into the Collins diffraction formula.
The latter one has the advantage of providing some analytical constraints which determine the validity of the proposed integral. 
In this sense, the evaluation of the focal field is restricted to a bounded region around the focus (for a given frequency).
As in the standard analysis, the inclusion of the focal region into the region of validity of the Debye approximation is guaranteed when the Fresnel number of the focusing geometry greatly exceeds unity.
Under the above assumptions, diffraction of purely-absorbing screens generates focal fields whose spectral components show symmetrical amplitude distributions around the focus.

In a second stage, the matrix-based Debye integral has been applied to investigate the spectral response of an apertured thin lens, either diffracting or refractive.
The position of the pupil aperture has a strong relevance in the Fresnel number, being the telecentric geometry the most appropriate design for the Debye representation.
Due to the longitudinal chromatic aberration, an angular dispersion effect is observable and may be important in the case of Fresnel lenses.

Going one step further, we have also examined the optical setup proposed by Lancis \textit{et al.} \cite{Lancis04} in the frame of the Debye representation.
This system shows a highly-dispersive numerical aperture, approaching the angular dispersion of an apertured Fresnel lens, but being free of longitudinal chromatic aberration.
The numerical aperture of the focusing system decreases with frequency, approximately following an inverse dependence.
The dispersive nature of diffraction is then compensated in the transverse focal plane, showing nearly frequency-independent Fraunhofer patterns.
As a collateral effect, a positive magnification of the focused field along the optical axis may be recognized.

In short, the present formalism leaves door open to the analysis and design of focused beams with variable angular dispersion.


\begin{thebibliography}{40}
\expandafter\ifx\csname natexlab\endcsname\relax\def\natexlab#1{#1}\fi
\expandafter\ifx\csname bibnamefont\endcsname\relax
  \def\bibnamefont#1{#1}\fi
\expandafter\ifx\csname bibfnamefont\endcsname\relax
  \def\bibfnamefont#1{#1}\fi
\expandafter\ifx\csname citenamefont\endcsname\relax
  \def\citenamefont#1{#1}\fi
\expandafter\ifx\csname url\endcsname\relax
  \def\url#1{\texttt{#1}}\fi
\expandafter\ifx\csname urlprefix\endcsname\relax\def\urlprefix{URL }\fi
\providecommand{\bibinfo}[2]{#2}
\providecommand{\eprint}[2][]{\url{#2}}

\bibitem[{\citenamefont{Corkum}(1993)}]{Corkum93}
\bibinfo{author}{\bibfnamefont{P.~B.} \bibnamefont{Corkum}},
  \bibinfo{journal}{Phys. Rev. Lett.} \textbf{\bibinfo{volume}{71}},
  \bibinfo{pages}{1994} (\bibinfo{year}{1993}).

\bibitem[{\citenamefont{Codling and Frasinski}(1993)}]{Codling93}
\bibinfo{author}{\bibfnamefont{K.}~\bibnamefont{Codling}} \bibnamefont{and}
  \bibinfo{author}{\bibfnamefont{L.~J.} \bibnamefont{Frasinski}},
  \bibinfo{journal}{J. Phys. B} \textbf{\bibinfo{volume}{26}},
  \bibinfo{pages}{783} (\bibinfo{year}{1993}).

\bibitem[{\citenamefont{Oron and Silberberg}(2005)}]{Oron05}
\bibinfo{author}{\bibfnamefont{D.}~\bibnamefont{Oron}} \bibnamefont{and}
  \bibinfo{author}{\bibfnamefont{Y.}~\bibnamefont{Silberberg}},
  \bibinfo{journal}{J. Opt. Soc. Am. B} \textbf{\bibinfo{volume}{22}},
  \bibinfo{pages}{2660} (\bibinfo{year}{2005}).

\bibitem[{\citenamefont{Bartels et~al.}(2000)\citenamefont{Bartels, Backus,
  Zeek, Misoguti, Vdovin, Christov, Murnane, and Kapteyn}}]{Bartels00}
\bibinfo{author}{\bibfnamefont{R.}~\bibnamefont{Bartels}},
  \bibinfo{author}{\bibfnamefont{S.}~\bibnamefont{Backus}},
  \bibinfo{author}{\bibfnamefont{E.}~\bibnamefont{Zeek}},
  \bibinfo{author}{\bibfnamefont{L.}~\bibnamefont{Misoguti}},
  \bibinfo{author}{\bibfnamefont{G.}~\bibnamefont{Vdovin}},
  \bibinfo{author}{\bibfnamefont{I.~P.} \bibnamefont{Christov}},
  \bibinfo{author}{\bibfnamefont{M.~M.} \bibnamefont{Murnane}},
  \bibnamefont{and} \bibinfo{author}{\bibfnamefont{H.~C.}
  \bibnamefont{Kapteyn}}, \bibinfo{journal}{Nature}
  \textbf{\bibinfo{volume}{406}}, \bibinfo{pages}{164} (\bibinfo{year}{2000}).

\bibitem[{\citenamefont{Dharmadhikari et~al.}(2004)\citenamefont{Dharmadhikari,
  Rajgara, Reddy, Sandhu, and Mathur}}]{Dharmadhikari04}
\bibinfo{author}{\bibfnamefont{A.}~\bibnamefont{Dharmadhikari}},
  \bibinfo{author}{\bibfnamefont{F.}~\bibnamefont{Rajgara}},
  \bibinfo{author}{\bibfnamefont{N.~C.} \bibnamefont{Reddy}},
  \bibinfo{author}{\bibfnamefont{A.}~\bibnamefont{Sandhu}}, \bibnamefont{and}
  \bibinfo{author}{\bibfnamefont{D.}~\bibnamefont{Mathur}},
  \bibinfo{journal}{Opt. Express} \textbf{\bibinfo{volume}{12}},
  \bibinfo{pages}{695} (\bibinfo{year}{2004}).

\bibitem[{\citenamefont{Wadsworth et~al.}(2002)\citenamefont{Wadsworth,
  Ortigosa-Blanch, Knight, Birks, Man, and Russell}}]{Wadsworth02}
\bibinfo{author}{\bibfnamefont{W.~J.} \bibnamefont{Wadsworth}},
  \bibinfo{author}{\bibfnamefont{A.}~\bibnamefont{Ortigosa-Blanch}},
  \bibinfo{author}{\bibfnamefont{J.~C.} \bibnamefont{Knight}},
  \bibinfo{author}{\bibfnamefont{T.~A.} \bibnamefont{Birks}},
  \bibinfo{author}{\bibfnamefont{T.-P.~M.} \bibnamefont{Man}},
  \bibnamefont{and} \bibinfo{author}{\bibfnamefont{P.~S.~J.}
  \bibnamefont{Russell}}, \bibinfo{journal}{J. Opt. Soc. Am. B}
  \textbf{\bibinfo{volume}{19}}, \bibinfo{pages}{2148} (\bibinfo{year}{2002}).

\bibitem[{\citenamefont{Konig}(2000)}]{Konig00}
\bibinfo{author}{\bibfnamefont{K.}~\bibnamefont{Konig}}, \bibinfo{journal}{J.
  Microsc. -Oxf.} \textbf{\bibinfo{volume}{200}}, \bibinfo{pages}{83}
  (\bibinfo{year}{2000}).

\bibitem[{\citenamefont{Barad et~al.}(1997)\citenamefont{Barad, Eisenberg,
  Horowitz, and Silberberg}}]{Barad97}
\bibinfo{author}{\bibfnamefont{Y.}~\bibnamefont{Barad}},
  \bibinfo{author}{\bibfnamefont{H.}~\bibnamefont{Eisenberg}},
  \bibinfo{author}{\bibfnamefont{M.}~\bibnamefont{Horowitz}}, \bibnamefont{and}
  \bibinfo{author}{\bibfnamefont{Y.}~\bibnamefont{Silberberg}},
  \bibinfo{journal}{Appl. Phys. Lett.} \textbf{\bibinfo{volume}{70}},
  \bibinfo{pages}{922} (\bibinfo{year}{1997}).

\bibitem[{\citenamefont{Cernusca et~al.}(1998)\citenamefont{Cernusca, Hofer,
  and Reider}}]{Cernusca98}
\bibinfo{author}{\bibfnamefont{M.}~\bibnamefont{Cernusca}},
  \bibinfo{author}{\bibfnamefont{M.}~\bibnamefont{Hofer}}, \bibnamefont{and}
  \bibinfo{author}{\bibfnamefont{G.~A.} \bibnamefont{Reider}},
  \bibinfo{journal}{J. Opt. Soc. Am. B} \textbf{\bibinfo{volume}{15}},
  \bibinfo{pages}{2476} (\bibinfo{year}{1998}).

\bibitem[{\citenamefont{Kempe et~al.}(1992)\citenamefont{Kempe, Stamm,
  Wilhelmi, and Rudolph}}]{Kempe92}
\bibinfo{author}{\bibfnamefont{M.}~\bibnamefont{Kempe}},
  \bibinfo{author}{\bibfnamefont{U.}~\bibnamefont{Stamm}},
  \bibinfo{author}{\bibfnamefont{B.}~\bibnamefont{Wilhelmi}}, \bibnamefont{and}
  \bibinfo{author}{\bibfnamefont{W.}~\bibnamefont{Rudolph}},
  \bibinfo{journal}{J. Opt. Soc. Am. B} \textbf{\bibinfo{volume}{9}},
  \bibinfo{pages}{1158} (\bibinfo{year}{1992}).

\bibitem[{\citenamefont{Kempe and Rudolph}(1993)}]{Kempe93}
\bibinfo{author}{\bibfnamefont{M.}~\bibnamefont{Kempe}} \bibnamefont{and}
  \bibinfo{author}{\bibfnamefont{W.}~\bibnamefont{Rudolph}},
  \bibinfo{journal}{Phys. Rev. A} \textbf{\bibinfo{volume}{48}},
  \bibinfo{pages}{4721} (\bibinfo{year}{1993}).

\bibitem[{\citenamefont{Zhu et~al.}(2005)\citenamefont{Zhu, Howe, Durst,
  Zipfel, and Xu}}]{Zhu05}
\bibinfo{author}{\bibfnamefont{G.}~\bibnamefont{Zhu}},
  \bibinfo{author}{\bibfnamefont{J.~V.} \bibnamefont{Howe}},
  \bibinfo{author}{\bibfnamefont{M.}~\bibnamefont{Durst}},
  \bibinfo{author}{\bibfnamefont{W.}~\bibnamefont{Zipfel}}, \bibnamefont{and}
  \bibinfo{author}{\bibfnamefont{C.}~\bibnamefont{Xu}}, \bibinfo{journal}{Opt.
  Express} \textbf{\bibinfo{volume}{13}}, \bibinfo{pages}{2153}
  (\bibinfo{year}{2005}).

\bibitem[{\citenamefont{Zeng et~al.}(2006)\citenamefont{Zeng, Lv, Zhan, Chen,
  Xiong, Luo, and Jacques}}]{Zeng06}
\bibinfo{author}{\bibfnamefont{S.~Q.} \bibnamefont{Zeng}},
  \bibinfo{author}{\bibfnamefont{X.}~\bibnamefont{Lv}},
  \bibinfo{author}{\bibfnamefont{C.}~\bibnamefont{Zhan}},
  \bibinfo{author}{\bibfnamefont{W.~R.} \bibnamefont{Chen}},
  \bibinfo{author}{\bibfnamefont{W.~H.} \bibnamefont{Xiong}},
  \bibinfo{author}{\bibfnamefont{Q.}~\bibnamefont{Luo}}, \bibnamefont{and}
  \bibinfo{author}{\bibfnamefont{S.~L.} \bibnamefont{Jacques}},
  \bibinfo{journal}{Opt. Lett.} \textbf{\bibinfo{volume}{31}},
  \bibinfo{pages}{1091} (\bibinfo{year}{2006}).

\bibitem[{\citenamefont{Amako et~al.}(2002)\citenamefont{Amako, Nagasaka, and
  Kazuhiro}}]{Amako02}
\bibinfo{author}{\bibfnamefont{J.}~\bibnamefont{Amako}},
  \bibinfo{author}{\bibfnamefont{K.}~\bibnamefont{Nagasaka}}, \bibnamefont{and}
  \bibinfo{author}{\bibfnamefont{N.}~\bibnamefont{Kazuhiro}},
  \bibinfo{journal}{Opt. Lett.} \textbf{\bibinfo{volume}{27}},
  \bibinfo{pages}{969} (\bibinfo{year}{2002}).

\bibitem[{\citenamefont{Li et~al.}(2005)\citenamefont{Li, Zhou, and Dai}}]{Li05}
\bibinfo{author}{\bibfnamefont{F.}~\bibnamefont{Li}},
  \bibinfo{author}{\bibfnamefont{C.}~\bibnamefont{Zhou}}, \bibnamefont{and}
  \bibinfo{author}{\bibfnamefont{E.} \bibnamefont{Dai}},
  \bibinfo{journal}{J. Opt. Soc. Am. A} \textbf{\bibinfo{volume}{22}},
  \bibinfo{pages}{767} (\bibinfo{year}{2005}).

\bibitem[{\citenamefont{Morris}(1981)}]{Morris81}
\bibinfo{author}{\bibfnamefont{G.~M.} \bibnamefont{Morris}},
  \bibinfo{journal}{Appl. Opt.} \textbf{\bibinfo{volume}{20}},
  \bibinfo{pages}{2017} (\bibinfo{year}{1981}).

\bibitem[{\citenamefont{Lancis et~al.}(1999)\citenamefont{Lancis, Tajahuerce, Andr\'es,
  M\'{\i}nguez-Vega, Fern\'andez-Alonso, and Climent}}]{Lancis99}
\bibinfo{author}{\bibfnamefont{J.}~\bibnamefont{Lancis}},
  \bibinfo{author}{\bibfnamefont{E.}~\bibnamefont{Tajahuerce}},
  \bibinfo{author}{\bibfnamefont{P.}~\bibnamefont{Andr\'es}}, 
  \bibinfo{author}{\bibfnamefont{G.}~\bibnamefont{M\'{\i}nguez-Vega}},
  \bibinfo{author}{\bibfnamefont{M.}~\bibnamefont{Fern\'andez-Alonso}},
  \bibnamefont{and}
  \bibinfo{author}{\bibfnamefont{V.}~\bibnamefont{Climent}},
  \bibinfo{journal}{Opt. Commun.} \textbf{\bibinfo{volume}{172}},
  \bibinfo{pages}{153} (\bibinfo{year}{1999}).

\bibitem[{\citenamefont{Lancis et~al.}(2004)\citenamefont{Lancis,
  M\'{\i}nguez-Vega, Tajahuerce, Climent, Andr\'es, and
  Caraquitena}}]{Lancis04}
\bibinfo{author}{\bibfnamefont{J.}~\bibnamefont{Lancis}},
  \bibinfo{author}{\bibfnamefont{G.}~\bibnamefont{M\'{\i}nguez-Vega}},
  \bibinfo{author}{\bibfnamefont{E.}~\bibnamefont{Tajahuerce}},
  \bibinfo{author}{\bibfnamefont{V.}~\bibnamefont{Climent}},
  \bibinfo{author}{\bibfnamefont{P.}~\bibnamefont{Andr\'es}}, \bibnamefont{and}
  \bibinfo{author}{\bibfnamefont{J.}~\bibnamefont{Caraquitena}},
  \bibinfo{journal}{J. Opt. Soc. Am. A} \textbf{\bibinfo{volume}{21}},
  \bibinfo{pages}{1875} (\bibinfo{year}{2004}).

\bibitem[{\citenamefont{Born and Wolf}(1999)}]{Born99}
\bibinfo{author}{\bibfnamefont{M.}~\bibnamefont{Born}} \bibnamefont{and}
  \bibinfo{author}{\bibfnamefont{E.}~\bibnamefont{Wolf}},
  \emph{\bibinfo{title}{Principles of Optics, Seventh (expanded) edition}}
  (\bibinfo{publisher}{Cambridge University Press}, \bibinfo{year}{1999}).

\bibitem[{\citenamefont{Stamnes}(1986)}]{Stamnes86}
\bibinfo{author}{\bibfnamefont{J.~J.} \bibnamefont{Stamnes}},
  \emph{\bibinfo{title}{Waves in Focal Regions}} (\bibinfo{publisher}{Adam
  Hilger}, \bibinfo{address}{Bristol and Boston}, \bibinfo{year}{1986}).

\bibitem[{\citenamefont{Collet and Wolf}(1980)}]{Collet80}
\bibinfo{author}{\bibfnamefont{E.}~\bibnamefont{Collet}} \bibnamefont{and}
  \bibinfo{author}{\bibfnamefont{E.}~\bibnamefont{Wolf}},
  \bibinfo{journal}{Opt. Lett.} \textbf{\bibinfo{volume}{5}},
  \bibinfo{pages}{264} (\bibinfo{year}{1980}).

\bibitem[{\citenamefont{Wolf and Li}(1981)}]{Wolf81}
\bibinfo{author}{\bibfnamefont{E.}~\bibnamefont{Wolf}} \bibnamefont{and}
  \bibinfo{author}{\bibfnamefont{Y.}~\bibnamefont{Li}}, \bibinfo{journal}{Opt.
  Commun.} \textbf{\bibinfo{volume}{39}}, \bibinfo{pages}{205}
  (\bibinfo{year}{1981}).

\bibitem[{\citenamefont{Pearce and Mittleman}(2002)}]{Pearce02}
\bibinfo{author}{\bibfnamefont{J.}~\bibnamefont{Pearce}} \bibnamefont{and}
  \bibinfo{author}{\bibfnamefont{D.}~\bibnamefont{Mittleman}},
  \bibinfo{journal}{Phys. Rev. E} \textbf{\bibinfo{volume}{66}},
  \bibinfo{pages}{056602} (\bibinfo{year}{2002}).

\bibitem[{\citenamefont{Gbur et~al.}(2002)\citenamefont{Gbur, Visser, and
  Wolf}}]{Gbur02}
\bibinfo{author}{\bibfnamefont{G.}~\bibnamefont{Gbur}},
  \bibinfo{author}{\bibfnamefont{T.~D.} \bibnamefont{Visser}},
  \bibnamefont{and} \bibinfo{author}{\bibfnamefont{E.}~\bibnamefont{Wolf}},
  \bibinfo{journal}{Phys. Rev. Lett.} \textbf{\bibinfo{volume}{88}},
  \bibinfo{pages}{013901} (\bibinfo{year}{2002}).

\bibitem[{\citenamefont{O'Hara}(2003)}]{Ohara98}
\bibinfo{author}{\bibfnamefont{J.~F.} \bibnamefont{O'Hara}}, Ph.D. thesis,
  \bibinfo{school}{Oklahoma State University} (\bibinfo{year}{2003}).

\bibitem[{\citenamefont{Zapata-Rodr\'{\i}guez}(2006{\natexlab{a}})}]{Zapata06}
\bibinfo{author}{\bibfnamefont{C.~J.} \bibnamefont{Zapata-Rodr\'{\i}guez}},
  \bibinfo{journal}{Opt. Commun.} \textbf{\bibinfo{volume}{257}},
  \bibinfo{pages}{9} (\bibinfo{year}{2006}{\natexlab{a}}).

\bibitem[{\citenamefont{Zapata-Rodr\'{\i}guez}(2006{\natexlab{b}})}]{Zapata06b}
\bibinfo{author}{\bibfnamefont{C.~J.} \bibnamefont{Zapata-Rodr\'{\i}guez}},
  \bibinfo{journal}{Opt. Commun.} 
(\bibinfo{year}{to be published}{\natexlab{b}}).

\bibitem[{\citenamefont{Collins}(1970)}]{Collins70}
\bibinfo{author}{\bibfnamefont{S.~A.} \bibnamefont{Collins}},
  \bibinfo{journal}{J. Opt. Soc. Am.} \textbf{\bibinfo{volume}{60}},
  \bibinfo{pages}{1168} (\bibinfo{year}{1970}).

\bibitem[{\citenamefont{Siegman}(1986)}]{Siegman86}
\bibinfo{author}{\bibfnamefont{A.~E.} \bibnamefont{Siegman}},
  \emph{\bibinfo{title}{Lasers}} (\bibinfo{publisher}{University Science
  Books}, \bibinfo{address}{Mill Valley}, \bibinfo{year}{1986}).

\bibitem[{\citenamefont{Yura and Hanson}(1987)}]{Yura87}
\bibinfo{author}{\bibfnamefont{H.~T.} \bibnamefont{Yura}} \bibnamefont{and}
  \bibinfo{author}{\bibfnamefont{S.~G.} \bibnamefont{Hanson}},
  \bibinfo{journal}{J. Opt. Soc. Am. A} \textbf{\bibinfo{volume}{4}},
  \bibinfo{pages}{1931} (\bibinfo{year}{1987}).

\bibitem[{\citenamefont{Gu}(2000)}]{Gu00}
\bibinfo{author}{\bibfnamefont{M.}~\bibnamefont{Gu}},
  \emph{\bibinfo{title}{Advanced Optical Imaging Theory}}
  (\bibinfo{publisher}{Heidelberg}, \bibinfo{address}{Springer},
  \bibinfo{year}{2000}).

\bibitem[{\citenamefont{Zapata-Rodr\'{\i}guez
  et~al.}(2000)\citenamefont{Zapata-Rodr\'{\i}guez, Andr\'es,
  Mart\'{\i}nez-Corral, and Mu{\~n}oz-Escriv\'a}}]{Zapata00}
\bibinfo{author}{\bibfnamefont{C.~J.} \bibnamefont{Zapata-Rodr\'{\i}guez}},
  \bibinfo{author}{\bibfnamefont{P.}~\bibnamefont{Andr\'es}},
  \bibinfo{author}{\bibfnamefont{M.}~\bibnamefont{Mart\'{\i}nez-Corral}},
  \bibnamefont{and}
  \bibinfo{author}{\bibfnamefont{L.}~\bibnamefont{Mu{\~n}oz-Escriv\'a}},
  \bibinfo{journal}{J. Opt. Soc. Am. A} \textbf{\bibinfo{volume}{17}},
  \bibinfo{pages}{1185} (\bibinfo{year}{2000}).

\bibitem{Sheppard88Martinez99} A collection of adapted pupil screens may be found in 
\bibinfo{author}{\bibfnamefont{C.~J.~R.} \bibnamefont{Sheppard}}
  \bibnamefont{and} \bibinfo{author}{\bibfnamefont{Z.~S.}
  \bibnamefont{Hegedus}}, \bibinfo{journal}{J. Opt. Soc. Am. A}
  \textbf{\bibinfo{volume}{5}}, \bibinfo{pages}{643} (\bibinfo{year}{1988}),
and
\bibinfo{author}{\bibfnamefont{M.}~\bibnamefont{Mart\'{\i}nez-Corral}},
  \bibinfo{author}{\bibfnamefont{P.}~\bibnamefont{Andr\'es}},
  \bibinfo{author}{\bibfnamefont{C.~J.} \bibnamefont{Zapata-Rodr\'{\i}guez}},
  \bibnamefont{and}
  \bibinfo{author}{\bibfnamefont{M.}~\bibnamefont{Kowalczyk}},
  \bibinfo{journal}{Opt. Commun.} \textbf{\bibinfo{volume}{165}},
  \bibinfo{pages}{267} (\bibinfo{year}{1999}).

\bibitem{Sellmeier} The Sellmeier coefficients for many common optical glasses can be found in the Schott Glass catalogue (See http://www.schott.com).

\bibitem[{\citenamefont{Levy et~al.}(2001)\citenamefont{Levy, Mendlovic, and
  Marom}}]{Levi01}
\bibinfo{author}{\bibfnamefont{U.}~\bibnamefont{Levy}},
  \bibinfo{author}{\bibfnamefont{D.}~\bibnamefont{Mendlovic}},
  \bibnamefont{and} \bibinfo{author}{\bibfnamefont{E.}~\bibnamefont{Marom}},
  \bibinfo{journal}{J. Opt. Soc. Am. A} \textbf{\bibinfo{volume}{18}},
  \bibinfo{pages}{86} (\bibinfo{year}{2001}).

\bibitem[{\citenamefont{Ashman and Gu}(2003)}]{Ashman03}
\bibinfo{author}{\bibfnamefont{R.}~\bibnamefont{Ashman}} \bibnamefont{and}
  \bibinfo{author}{\bibfnamefont{M.}~\bibnamefont{Gu}}, \bibinfo{journal}{Appl.
  Opt.} \textbf{\bibinfo{volume}{42}}, \bibinfo{pages}{1852}
  (\bibinfo{year}{2003}).

\bibitem[{\citenamefont{Zapata-Rodr\'{\i}guez}(2006{\natexlab{c}})}]{Zapata06c}
\bibinfo{author}{\bibfnamefont{C.~J.} \bibnamefont{Zapata-Rodr\'{\i}guez}},
  \bibinfo{journal}{J. Opt. Soc. Am. A} 
(\bibinfo{year}{to be published}{\natexlab{c}}).

\bibitem[{\citenamefont{Moreno et~al.}(1997)\citenamefont{Moreno, Rom\'an, and
  Salgueiro}}]{Moreno97}
\bibinfo{author}{\bibfnamefont{V.}~\bibnamefont{Moreno}},
  \bibinfo{author}{\bibfnamefont{J.~F.} \bibnamefont{Rom\'an}},
  \bibnamefont{and} \bibinfo{author}{\bibfnamefont{J.~R.}
  \bibnamefont{Salgueiro}}, \bibinfo{journal}{Am. J. Phys.}
  \textbf{\bibinfo{volume}{65}}, \bibinfo{pages}{556} (\bibinfo{year}{1997}).

\bibitem[{\citenamefont{Fischer and Tadic}(2000)}]{Fischer00b}
\bibinfo{author}{\bibfnamefont{R.~F.} \bibnamefont{Fischer}} \bibnamefont{and}
  \bibinfo{author}{\bibfnamefont{B.}~\bibnamefont{Tadic}},
  \emph{\bibinfo{title}{Optical system design}}
  (\bibinfo{publisher}{McGraw-Hill Professional}, \bibinfo{year}{2000}).

\bibitem[{\citenamefont{Ibragimov}(1995)}]{Ibragimov95}
\bibinfo{author}{\bibfnamefont{E.}~\bibnamefont{Ibragimov}},
  \bibinfo{journal}{Appl. Opt.} \textbf{\bibinfo{volume}{34}},
  \bibinfo{pages}{7280} (\bibinfo{year}{1995}).

\end{thebibliography}








\end{document}